\documentclass[a4paper,conference]{IEEEtran}
\IEEEoverridecommandlockouts
\usepackage{cite}
\usepackage{amsmath,amssymb,amsfonts}
\usepackage{algorithmic}
\usepackage{graphicx}
\usepackage{textcomp}
\usepackage{xcolor}
\usepackage{subfiles}
\usepackage{balance}
\usepackage{multirow}
\usepackage{subfigure}

\def\BibTeX{{\rm B\kern-.05em{\sc i\kern-.025em b}\kern-.08em
    T\kern-.1667em\lower.7ex\hbox{E}\kern-.125emX}}

\usepackage{makecell}
\begin{document}

\title{Audio-based Near-Duplicate Video Retrieval with Audio Similarity Learning}

\author{\\Pavlos Avgoustinakis$^1$ \quad Giorgos Kordopatis-Zilos$^1$ \quad Symeon Papadopoulos$^1$ \\ Andreas L. Symeonidis$^2$ \quad Ioannis Kompatsiaris$^1$ \vspace{0.1cm}\\
$^1$Information Technologies Institute, CERTH, Thessaloniki, Greece \\
$^2$Electrical and Computer Engineering Dept., Aristotle University of Thessaloniki, Greece \\
{\tt\small  \{pavgoust,georgekordopatis,papadop,ikom\}@iti.gr \quad asymeon@eng.auth.gr}\\}

\maketitle

\begin{abstract}
In this work, we address the problem of audio-based near-duplicate video retrieval. We propose the Audio Similarity Learning (AuSiL) approach that effectively captures temporal patterns of audio similarity between video pairs. For the robust similarity calculation between two videos, we first extract representative audio-based video descriptors by leveraging transfer learning based on a Convolutional Neural Network (CNN) trained on a large scale dataset of audio events, and then we calculate the similarity matrix derived from the pairwise similarity of these descriptors. The similarity matrix is subsequently fed to a CNN network that captures the temporal structures existing within its content. We train our network following a triplet generation process and optimizing the triplet loss function. To evaluate the effectiveness of the proposed approach, we have manually annotated two publicly available video datasets based on the audio duplicity between their videos. The proposed approach achieves very competitive results compared to three state-of-the-art methods. Also, unlike the competing methods, it is very robust to the retrieval of audio duplicates generated with speed transformations.
\end{abstract}


\section{Introduction}
The increasing availability of affordable recording devices and the rapid growth of online video platforms, such as YouTube\footnote{https://www.youtube.com/} and TikTok\footnote{https://www.tiktok.com/}, has led to the explosive increase of the volume of video data. In tandem, there is an overwhelming growth of duplicate video content shared online, e.g., by users who re-post processed or edited versions of original videos on social media platforms. This makes near-duplicate video retrieval (NDVR) a research topic of growing importance in the last few years. In this paper, we address a special case of the NDVR problem, which is the retrieval of videos that are duplicate in terms of their audio content. The manipulated audio content of videos may have undergone various transformations, i.e., mp3 compression, bandwidth limitation, or mix with speech. We will refer to this instance of the problem as Duplicate Audio Video Retrieval (DAVR).

Although many NDVR methods exist that exploit the visual content of videos to perform retrieval, to the best of our knowledge, no method that addresses the DAVR problem has been proposed. Nevertheless, there are approaches in the literature that tackle similar audio-based retrieval problems, such as Content-Based Copy Detection (CBCD). Such methods usually extract audio fingerprints using handcrafted processes. However, no CBCD method employs deep learning techniques, which is a common practice in the corresponding visual-based version of the problem. Moreover, transfer learning is widely used in the computer vision field because of the availability of large datasets such as \textit{ImageNet} \cite{deng2009imagenet}. In the case of audio, transfer learning has been less explored until recently due to the unavailability of similar large-scale datasets. Additionally, there is no publicly available video dataset with user-generated content that is annotated based on audio duplicity to evaluate DAVR methods.

Recently, some methods have been proposed that can be employed in order to address the problem of DAVR. Kumar et al. \cite{kumar2018knowledge} proposed a method to effectively transfer knowledge from a sound event classification model based on a Convolutional Neural Network (CNN). They trained their model on \textit{AudioSet} \cite{audioset}, a recently released weakly labeled dataset with sound events. The knowledge transfer capability of the pretrained CNN was evaluated on several audio recognition tasks and was found to generalize well, reaching human-level accuracy on environmental sound classification. Moreover, Kordopatis et al. \cite{kordopatis2019visil} recently introduced ViSiL, a video similarity learning architecture that exploits spatio-temporal relations of the visual content to calculate the similarity between pairs of videos. It is a CNN-based approach trained to compute video-to-video similarity from frame-to-frame similarity matrices, considering intra- and inter-frame relations. The proposed method was evaluated on several visual-based video retrieval problems exceeding the state-of-the-art. 

Our motivation in this paper is to build an audio-based approach that employs transfer learning and video similarity learning in order to address the DAVR problem. Additionally, due to the lack of a suitable dataset for the evaluation of such approaches, our goal is to compose annotated corpora that serve as evaluation testbeds for DAVR. To this end, we propose AuSiL, an audio similarity learning approach. In the proposed approach, we extract features from the activations of the intermediate convolutional layers of the pretrained CNN architecture\cite{kumar2018knowledge} that is fed with the Mel-spectrograms of the audio signals of videos. In that way, we extract compact audio descriptors for the video frames. The audio-based video representations are further refined by applying PCA whitening and attention weighting. To compute the similarity between video pairs, we first calculate the similarity matrix that contains the pairwise similarity scores between the audio descriptors. Then, we propagate the similarity matrix to a CNN network that captures the temporal similarity patterns and calculates the final similarity between the two videos. Furthermore, we develop a triplet generation process to form video triplets, and we train our model by optimizing the triplet loss function. To cover the benchmarking needs of the DAVR task, we have manually annotated the publicly available FIVR-200K \cite{kordopatis2019fivr} and SVD \cite{jiang2019svd} datasets by labeling videos that share duplicate audio segments with the queries. The proposed approach is compared against three competing methods. It demonstrates very competitive performance and proves to be very robust to the retrieval of audio duplicates generated with speed transformations, in contrast to the competing approaches.

\section{Related Work}
\label{related}

In this section, we briefly discuss several audio-based methods proposed for the CBCD problem, which is closely related to the DAVR. Typical CBCD methods consist of two parts: i) a process for the extraction of fingerprints that encode signal properties derived from the audio channels of videos, and ii) a search algorithm that calculates the similarity between the videos in a database and a given query video based on the extracted fingerprints. Additionally, we present several works that exploit transfer learning on audio-based problems.

A large variety of audio descriptors have been proposed in the literature. Roopalakshmi et al. \cite{roopalakshmi2011novel} proposed a video copy detection method based on audio fingerprints composed by the Mel-Frequency Cepstral Coefficients (MFCC) features and four spectral descriptors, reduced based on PCA. Jegou et al. \cite{jegou2012babaz} extracted features for short-term time windows based on 64 filter banks. The audio descriptors are created from the concatenation of the features of three successive time windows, resulting in a single descriptor of 192 dimensions that represents 83 ms of the audio signal. Another popular audio descriptor is the band energy difference \cite{haitsma2002highly,saracoglu2009content,wang2012contented}. Haitsma et al. \cite{haitsma2002highly} generated fingerprints for short term time windows, based on the monotonicity between 33 successive frequency sub-bands, resulting in a 32 bits hash descriptor. Saracoglu et al. \cite{saracoglu2009content} used energy differences between 16 sub-bands in order to reduce search time. Wang et al. \cite{wang2012contented} expanded this method by computing the differences between all sub-bands, not just successive ones, and choosing a subset that contains the most representative differences. One of the most popular audio fingerprints is proposed by the Shazam system \cite{wang2003industrial}. It generates binary audio descriptors by encoding the relations between two spectral peaks. To make the system more robust, Anguera et al. \cite{anguera2012mask} proposed an approach that selects salient points of the Mel-filtered spectrogram and then applies a mask centered at each of them, to define regions of interest. The audio fingerprints are encoded by comparing the energy of the regions. Ouali et al. \cite{ouali2014robust, ouali2015efficient, ouali2016fast} extracted audio descriptors by producing various versions of the spectrogram matrix of the audio signal, using values based on the average of spectral values for thresholding, resulting in 2-D binary images. They proposed two different schemes for the extraction of audio descriptors. In \cite{ouali2014robust}, the binary image is divided into horizontal and vertical slides. The fingerprint is composed of the sum of the elements of each slide. In \cite{ouali2015efficient, ouali2016fast}, the binary image is divided into tiles. The fingerprint is generated based on the positions of the tiles with the highest sum in the image.

Furthermore, many algorithms have been proposed for the calculation of the similarity between videos. To search the audio fingerprints in the database, the method in \cite{roopalakshmi2011novel} calculates the similarity between fingerprints, using weighted L2-Euclidean distance, while in \cite{jegou2012babaz} the similarity is estimated by exploiting the reciprocal nearest neighbors. In \cite{haitsma2002highly}, various sub-fingerprints are produced by altering the most unreliable bits of the original fingerprint in order to calculate the bit error rate between the sub-fingerprints of the audio descriptors of a query and reference video.
In \cite{saracoglu2009content, wang2012contented}, a voting system is employed that counts votes for the equal time differences of the matching fingerprints between a query and a reference video. The reference sequence with the highest vote count is regarded as a match. In the cases of binary images, every fingerprint of the query video is linked with the nearest neighbor fingerprint of the reference \cite{ouali2014robust,ouali2015efficient,ouali2016fast}. To quantify the distance between fingerprints, the authors employed the Manhattan distance in \cite{ouali2014robust} and the total number of coexisting positions in \cite{ouali2015efficient, ouali2016fast}. Then, the query shifts over the reference, and for each alignment, the number of matching query frames with their nearest neighbor is counted. The similarity between two compared videos is then computed according to the reference segment with the highest count. 


Yet, none of the related works in the CBCD field have experimented with features extracted from deep learning networks, a practice that has wide success in visual-based retrieval problems. In this paper, we evaluate the application of such features extracted from a CNN-based architecture proposed for transfer learning \cite{kumar2018knowledge}. Additionally, the proposed solutions for similarity calculation cannot capture a large variety of temporal similarity patterns due to their rigid aggregation approach. Therefore, to tackle this limitation, we build a similarity learning network to robustly compute the similarity between videos. For comparing our method with related works, we have reimplemented the \cite{ouali2014robust, ouali2016fast} approaches, because they reported competitive results, outperforming prior methods. We also compare against the Dejavu open-source framework \cite{dejavu2013}, which reimplements the popular Shazam system \cite{wang2003industrial}.

\section{Proposed Method}

The proposed system comprises two parts, the extraction of representative audio descriptors and the audio similarity calculation between pairs of video. First, we extract features from the intermediate convolutional layers of a CNN, which takes as input time segments of the audio spectrogram. Then, the extracted features are PCA whitened and weighted based on an attention mechanism. To estimate the similarity between videos, a similarity matrix with the pairwise segment similarities of two compared videos is propagated to a similarity learning CNN to capture the temporal patterns. The final similarity score is computed based on the Chamfer Similarity (CS) of the network's output. The model is trained using carefully selected triplets of video from a training dataset based on a triplet loss scheme.

\begin{figure*}
    \centering
    \includegraphics[width=17.7cm]{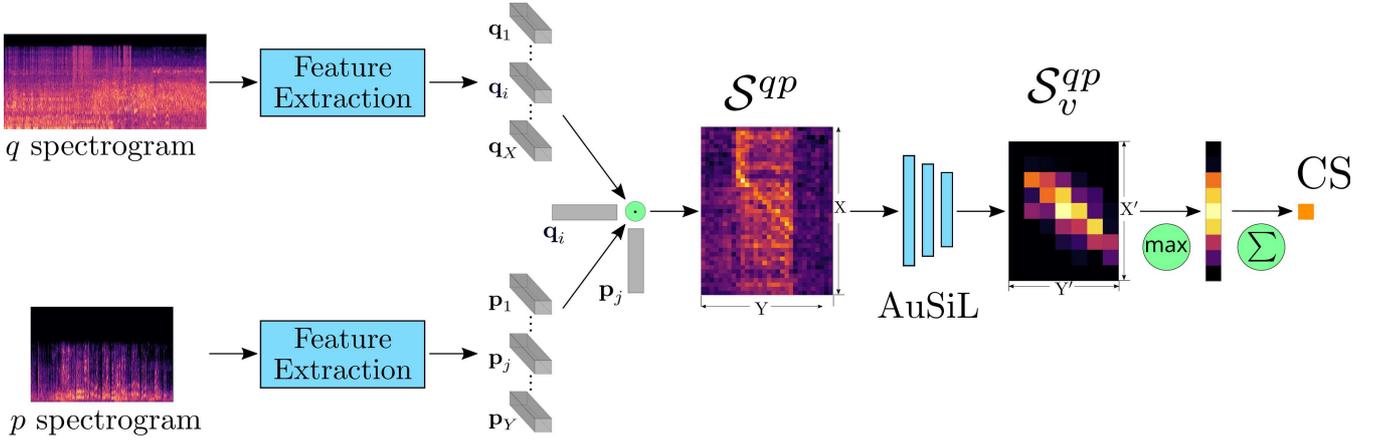}
    \caption{Similarity calculation process of the proposed architecture. The spectrogram of each video is provided to the feature extraction process, where feature vectors are extracted for each audio segment. Then, a similarity matrix is generated from the dot product between the feature vectors of the two videos. The generated matrix is provided to AuSiL CNN to capture the temporal patterns of the segment-level within-video similarities. The final similarity score is aggregated by applying Chamfer Similarity.}
    \label{pipeline}
    \vspace{-0.2cm}
\end{figure*}

\subsection{Feature Extraction}
\label{sec:feature_extraction}

To generate audio-based descriptors, we first extract the Mel-filtered spectrogram from the audio of videos. All audio signals are resampled at $44.1$ kHz sampling frequency. For the spectrogram generation, we use 128 Mel-bands and a window size of $23$ ms (1024 samples at $44.1$ kHz) with an overlap of $11.5$ ms (512 hop size). The generated spectrograms are divided into overlapping time segments of $2$ seconds with $t$ seconds time step. We consider $t$ as a system hyperparameter and we measure its effect on Section \ref{abbliation_sec}.

\begin{table}[t]
\centering
\caption{Number of features extracted from the intermediate convolutional layers of each block of the CNN, by applying MAC. These features compose the final 2,528-dimensions feature vector.}
\scalebox{1.}{
 \begin{tabular}{|c | c|} 
 \hline
 \textbf{CNN Block} & \textbf{Filter Size}  \\ 
 \hline\hline
 B1 & $16+16=32$ \\ 
 \hline
 B2 & $32+32=64$ \\ 
 \hline
 B3 & $64+64=128$ \\ 
 \hline
 B4 & $128+128=256$ \\ 
 \hline
 B5 & $256+256=512$ \\ 
 \hline
 B6 & $512$ \\ 
 \hline
 F1 & $1,024$ \\ 
 \hline\hline
 \textbf{Total} & $\textbf{2,528}$ \\ 
 \hline
\end{tabular}
}
\label{feature_composition}
\end{table}

Further, we feed the generated spectrogram segments to a feature extraction CNN designed for transfer learning, proposed by Kumar et al. \cite{kumar2018knowledge}. The CNN model is trained on the large-scale AudioSet \cite{audioset} dataset, consisting of approximately 2.1 million weakly labeled videos from YouTube with 527 audio event classes. The backbone CNN contains seven blocks, i.e., blocks B1-B6 and block F1. Each of B1-B5 blocks consists of two convolutional layers followed by a max-pooling layer. B6 consists of one convolutional layer, followed by max-pooling, and F1 consists of one convolutional layer. Batch normalization and a ReLU activation function are applied on the output of each convolutional layer.

To extract compact audio representations, we apply Maximum Activation of Convolution (MAC) on the activations of the intermediate convolutional layers of the feature extraction CNN model \cite{tolias2015particular, kordopatis2017a}. Given a CNN architecture with K convolutional layers, MAC generates K feature vectors $\textbf{h}^k\in\mathbb{R}^{C_k}$, where $C_k$ the number of channels of the $k^{th}$ convolutional layer. The extracted vectors are concatenated in a single feature vector $\textbf{h} \in \mathbb{R}^{C}$, where $C = C_1 + C_2 + ... + C_K$. We applied MAC on the intermediate layers of parts B1-B6 and F1 of the CNN. The dimensionality of the concatenated feature vector amounts to $2,528$ dimensions. Table \ref{feature_composition} presents the dimensionality of the feature vectors extracted from each block of the CNN. Then, we apply PCA whitening\cite{jegou2012negative} to decorrelate the feature vectors. The feature vectors are $\ell^2$-normalized before and after the concatenation and also, after the PCA whitening.

Applying $\ell^2$-normalization on the extracted feature vectors results in all audio segments having equal contribution to the similarity calculation. This could mean that, for instance, a silent segment would have the same impact as a segment with rich audio content. To overcome this issue, we employ a self-attention mechanism \cite{yang2016hierarchical} to weigh the audio segments based on their captured information. Given a feature vector $\textbf{h}$, we use a context vector $\textbf{u}$ to measure its importance. A weight score $a$ is derived by calculating the dot product between the feature vector $\textbf{h}$ and the context vector $\textbf{u}$. The resulting weight score will be in the range $[-1, 1]$ since all vectors have unit norm. However, to avoid the direction change of the feature vectors, we rescale the calculated weight scores in the $[0, 1]$ range, by dividing $a$ by $2$ and adding $0.5$. The weighting procedure is formulated in Equation \ref{attention}.

\begin{equation}
\begin{aligned}
    a &= \textbf{u}^T \textbf{h} \\
    \textbf{h}' &= (a/2 + 0.5) \textbf{h}
\end{aligned}
\label{attention}
\end{equation}

\begin{table}[t]
    \caption{Architecture of the proposed network. We assume that the similarity matrix of two videos with a total number of $X$ and $Y$ audio segments is provided as input.}
    \centering
    \scalebox{1.}{
    \begin{tabular}{|c|c|c|c|}
    \hline
    \textbf{Type} & \textbf{Kernel size / stride} & \textbf{Output size} & \textbf{Activ.} \\ \hline\hline
        \textbf{Conv} & $3\times3$ / $1$ & $X\times Y \times 32$ & ReLU \\ \hline
        \textbf{Max-Pool} & $2\times2$ / $2$ & $X/2 \times Y/2 \times 32$ & - \\ \hline
        \textbf{Conv} & $3\times3$ / $1$ & $X/2 \times Y/2 \times 64$ & ReLU \\ \hline
        \textbf{Max-Pool} & $2\times2$ / $2$ & $X/4 \times Y/4 \times 64$ & - \\ \hline
        \textbf{Conv} & $3\times3$ / $1$ & $X/4 \times Y/4 \times 128$ & ReLU \\ \hline
        \textbf{Conv} & $1\times1$ / $1$ & $X/4\times Y/4 \times 1$ & - \\ \hline
    \end{tabular}
    }
    \label{visilCNN}
\end{table}

\subsection{Similarity calculation}
To calculate video similarity, we first calculate the pairwise similarity matrix that contains the similarity scores between the audio feature vectors of two compared videos. More specifically, given two videos $q$, $p$, with $X$ and $Y$ audio segments respectively, we apply dot product between the feature vectors of the corresponding video descriptors $Q \in \mathbb{R}^{X\times{C}}$ and $P \in \mathbb{R}^{Y\times{C}}$, where $C$ is the dimensionality of feature vectors. This process produces a matrix $\mathcal{S}^{qp} \in\mathbb{R}^{X\times{Y}}$ containing the pairwise similarities between all vectors of the two videos, and can be formulated as a matrix multiplication in Equation \ref{sim_matrix}.
\begin{equation}
    \mathcal{S}^{qp} =  Q \cdot P^\top
    \label{sim_matrix}
\end{equation}
Then, the generated similarity matrix $S^{qp}$ is provided to a four-layer similarity learning CNN network \cite{kordopatis2019visil}. The network has the capability of capturing the temporal patterns of segment-level within-video similarities. The architecture of the proposed CNN is displayed in Table \ref{visilCNN}. Figure \ref{pipeline} depicts a visual example of the input and the output of the AuSiL CNN. The network can detect temporal patterns and assign high similarity scores in the corresponding segments, i.e., the diagonal part existing in the center of the similarity matrix. At the same time, the noise in the input matrix, introduced by the similarity calculation process, has been significantly reduced in the output. Next, we apply the \textit{hard tanh} activation function on the network output values to clip them in range $[-1,1]$. The final similarity score is derived by applying Chamfer Similarity (CS), which is formulated as a max operation followed by a mean operation, as in Equation \ref{chamfer}.
\begin{equation}
    \text{CS} (q, p) = \frac{1}{X'} \sum_{i=1}^{X'} \max_{j \in [1, Y']} \text{Htanh}(\mathcal{S}_{\upsilon}^{qp}(i,j)),
    \label{chamfer}
\end{equation}
where $\mathcal{S}_{\upsilon}^{qp} \in \mathbb{R}^{X'\times{Y'}}$ is the output of the CNN network and Htanh indicates the element-wise hard tanh function.


\subsection{Training process}
\label{sec:training_process}

Ideally, the video similarity score that derives from Equation \ref{chamfer} should be higher for videos that are relevant and lower for irrelevant ones. Therefore, we train our network by organising the training dataset in video triplets $(\upsilon, \upsilon ^+, \upsilon ^-)$, where $\upsilon, \upsilon ^+, \upsilon ^-$ stand for an anchor, a positive (relevant) and a negative (irrelevant) video respectively. For this purpose, we use the \textit{triplet loss} function \cite{schroff2015facenet}, as formulated in Equation \ref{triplet_loss}.
\begin{equation}
    \mathcal{L}_{tr} = \max \{0, \text{CS}(\upsilon,\upsilon^-) - \text{CS}(\upsilon,\upsilon^+) + \gamma \},
    \label{triplet_loss}
\end{equation}
where $\gamma$ is a margin parameter. Triplet loss forces the network to assign higher similarity scores to relevant pairs of video and lower scores to irrelevant ones. Additionally, we employ the similarity regularization loss described in \cite{kordopatis2019visil}, since it provides significant performance improvement. This loss function penalizes the network activations that are out of the clipping range of the \textit{hard tanh} activation function, as in Equation \ref{regularization}.
\begin{equation}
\begin{split}
    \mathcal{L}_{reg}=\sum_{i=1}^{X'} \sum_{j=1}^{Y'} |\max \{ 0, \mathcal{S}_\upsilon^{qp}(i,j)-1 \}| \\ 
    + |\min \{0, \mathcal{S}_\upsilon^{qp}(i,j)+1 \}|
    \label{regularization}
\end{split}
\end{equation}

The total loss function is defined in Equation \ref{eq:total_loss}.

\begin{equation}
    \mathcal{L} = \mathcal{L}_{tr} + r \cdot \mathcal{L}_{reg},
\label{eq:total_loss}
\end{equation}
where $r$ is a hyperparameter that determines the contribution of the similarity regularization to the total loss.

Training the architecture described above requires the organisation of the dataset used for training in video triplets. So, we extract pairs of videos with related audio content, to serve as anchor-positive pairs during training. Due to the unavailability of datasets with ground truth annotations in terms of audio content, we extract the positive pairs from a dataset with visual annotations. The videos that have not been labeled as positives are considered negatives. From all positive pairs in terms of visual content, we select only the ones whose global audio feature vectors' distance is smaller than a certain value. The global audio feature vectors of videos result from the application of global average pooling on the concatenated feature vectors (Section \ref{sec:feature_extraction}). The upper threshold value was empirically set to $0.175$. We then create video triplets based on the positive pairs by selecting videos that are \textit{hard negative} examples. More precisely, we select all the anchor-negative pairs whose Euclidean distance in the feature space is less than the distance between the anchor-positive pair plus a margin value $d$, i.e., $D(\upsilon,\upsilon^-)<D(\upsilon,\upsilon^+) + d$, where $D(\cdot, \cdot)$ indicates the Euclidean distance between two arbitrary videos. Value $d$ was empirically set to $0.15$.

\section{Evaluation Setup}

\subsection{Datasets}
\label{sec:datasets}

We employ the \textbf{VCDB} (Video Copy DataBase) \cite{jiang2014vcdb} to train our AuSiL network. This consists of videos collected from popular video platforms (YouTube and Metacafe) and has been compiled and annotated for the problem of partial copy detection. It contains $528$ videos with $9,236$ copied segments in the core set, and $100,000$ distractor videos in the background set. We use the videos in the core set to form the anchor-positive pairs, and we draw negatives from the background set. A total of 5.8 million triplets is formed from the triplet selection process.

\begin{table}[t]
  \centering
  \caption {Annotation labels of the FIVR-200K dataset along with their abbreviations and definitions.}
  \begin{tabular}{|m{1.9cm}|l|m{4.7cm}|}
      \hline
      \textbf{Label}   & \textbf{Abb.} & \textbf{Definition}       \\ \hline
       Near-Duplicate       & ND &  Videos that contain only duplicate scenes with the query  \\ \hline
       Duplicate Scene      & DS &  Videos that contain at least one duplicate scene with the query \\ \hline
       Complementary Scene  & CS &  Videos that depict the same incident moments with the query, but from a different viewpoint \\ \hline
       Incident Scene       & IS &  Videos that depict the same incident with the query, but has no temporal overlap     \\ \hline
    \end{tabular}
    \label{tab:transformation}
\end{table}

To build an evaluation corpus that simulates DAVR, we employ the \textbf{FIVR-200K} \cite{kordopatis2019fivr} dataset\footnote{http://ndd.iti.gr/fivr/}, which was originally composed for the problem of Fine-grained Incident Video Retrieval (FIVR). It contains 225,960 videos and 100 video queries collected from YouTube based on the major news events from recent years. Table \ref{tab:transformation} depicts the annotation labels used in the FIVR-200K, along with their definitions. For the simulation of the DAVR problem, we have set the following annotation procedure. We first select the queries that are suitable for the simulation of the DAVR problem;  we excluded 24 queries that were completely silent or noisy,  resulting in a set of 76 out of 100 queries. For each of them, we manually annotate the videos with ND, DS, and CS labels according to their audio duplicity with the query. The videos that share audio content with the queries are labeled with the Duplicate Audio (DA) label. In total, we validate 9,345 videos, from which 3,392 are labeled as DA. From this point on, we will refer to this audio-based annotated dataset as FIVR-200K\textsubscript{$\alpha$}. Also, for quick comparisons of the different variants of our proposed approach, we sample a subset of the original dataset, which we call \textbf{FIVR-5K\textsubscript{$\alpha$}}. For its composition, we first randomly select 50 queries, and then for each one, we randomly draw the 35\% of the videos labeled as DA. To make retrieval more challenging, we also add 5,000 distractor videos that are not related to the queries.

To build our second evaluation corpus, we employ the \textbf{SVD} (Short Video Dataset) \cite{jiang2019svd} dataset\footnote{https://svdbase.github.io/} that has been composed for the NDVR problem. The dataset consists of over 500,000 short videos collected from a large video platform (TikTok) and includes 1,206 query videos, 34,020 labeled videos, and 526,787 unlabeled videos that are likely not related to the queries. However, due to TikTok's nature, we empirically found that a large number of audio duplicates exist in the unlabeled set. Therefore, for the composition of an evaluation set that simulates DAVR, we consider only the videos in the labeled set of the SVD dataset. The annotation of the dataset is performed according to the following procedure. We first acquire all the query-candidate video pairs that have been labeled as positives by the original authors of the dataset, and we annotate the video pairs that share audio content. At the end of this process, we discard all queries with no video pairs annotated as positives, resulting in a query set of 167 out of 206 queries. To find potential audio duplicates that are not included in the labeled set, we manually annotate all query-candidate pairs that have not been labeled and have similarity greater than $0.4$. To compute the similarity scores, we follow the process described in Section \ref{sec:feature_extraction} to extract global feature vectors, and then use the dot product to measure similarity. Based on the described process, we composed an evaluation dataset consisting of 6,118 videos, 167 queries with 1,492 video pairs labeled as audio duplicates. From this point on, we will refer to this audio-based annotated dataset as SVD\textsubscript{$\alpha$}.

\subsection{Evaluation metrics}
\label{metrics}
For the evaluation of retrieval performance, we utilize \textit{mean average precision} (mAP), which quantifies the ranking of the database items given a query and thus is widely used as a measure for retrieval performance. For the computation of the mAP, we calculate the average precision (AP) for every video query, according to Equation \ref{eq:AP}.
\begin{equation}
    {AP} = \frac{1}{n} \sum_{i=1}^{n} \frac{i}{r_i}
\label{eq:AP}
\end{equation}
where $n$ is the number of relevant videos to the query and $r_i$ is the rank, based on the similarity score, of the $i$-th retrieved relevant video. The mAP is calculated by averaging the AP scores of all queries.

Additionally, to gain a better understanding about the methods' performance, we employ the interpolated precision-recall (PR) curve, which shows the trade-off between precision and recall for different thresholds.

\subsection{Implementation Details}
\label{sec:impl}

To train the network, we employ the Adam optimizer \cite{kingma2014adam} with learning rate $l=10^{-3}$ and regularization parameter $r=0.1$. Also, we consider as the default values of the hyperparameters $t=1s$ and $\gamma=1$. The parameters of PCA whitening are learned from a corpus of one million feature vectors sampled from the VCDB dataset. All experiments were conducted on a machine with Intel Xeon @2.10 GHz CPU and an Nvidia GTX1070 GPU. We trained the network for about 30 hours until convergence. For videos with average duration 100 s, the proposed system needs, on average, 100 ms for feature extraction per video, and 3 ms for the similarity calculation between a video pair.

\section{Experiments and Results}
\label{sec:experiments}

In this section, we present an ablation study by examining different configurations of the proposed approach (Section \ref{abbliation_sec}). Also, we compare AuSiL against three methods from the literature on the DAVR problem (Section \ref{DAVR_comparison_sec}). We evaluate the proposed approach to retrieval settings where audio speed transformations have been applied to the query videos (Section \ref{sec:speed_transform}). Lastly, we report results on the more challenging settings of three visual-based video retrieval tasks (Section \ref{visua_comp_sec}).

\subsection{Ablation study}
\label{abbliation_sec}

Initially, we study the impact of time step $t$ on the performance of AuSiL on the subset FIVR-5K\textsubscript{$\alpha$} and SVD\textsubscript{$\alpha$}. Table \ref{time_step_study} illustrates the mAP of the proposed method for different time step values. The time step appears to have a detrimental impact on the system's performance on SVD\textsubscript{$\alpha$}. The smaller time step values report clearly better results compared to the larger ones. Instead, this is not the case for FIVR-5K\textsubscript{$\alpha$}, where the selection of the time step seems to have limited impact on the system's performance. A possible explanation for this could be that SVD mainly consists of short duration videos, i.e., 17 seconds duration on average, unlike FIVR-200K, where the average duration is 117 seconds. Also, using smaller time step values generates larger similarity matrices with richer temporal patterns captured by the AuSiL, leading to more accurate similarity calculation. For the rest sections, we use a time step of 125 ms on the SVD dataset, and a time step of 1 s for all the others.

\begin{table}[htbp]
\caption{mAP comparison for various time steps (ms) on FIVR-5K\textsubscript{$\alpha$} and SVD\textsubscript{$\alpha$}}
\begin{center}
\begin{tabular}{|c||c|c|}
\hline

\textbf{Time step}  & \textbf{FIVR-5K\textsubscript{$\alpha$}} & \textbf{SVD\textsubscript{$\alpha$}} \\

\hline\hline

 1000 & $\textbf{0.794}$ & $0.903$\\
\hline
 500  & $0.789$ & $0.915$\\
\hline
 250  & $0.787$ & $0.928$\\
\hline
 125  & $0.790$ & $\textbf{0.940}$\\
\hline

\end{tabular}
\label{time_step_study}
\end{center}
\end{table}

We also examine the contribution of each AuSiL component. Table \ref{parts_study} shows the results on FIVR-5K\textsubscript{$\alpha$} and SVD\textsubscript{$\alpha$}, first using only the video features extracted from the feature extraction CNN and then adding PCA whitening, the attention mechanism, and the similarity learning CNN. The attention mechanism in every run is trained based on the main training process. Performance improves as individual components are added to the system. The application of PCA whitening has the most significant impact on the network's performance, with $0.084$ and $0.041$ mAP on FIVR-5K\textsubscript{$\alpha$} and SVD\textsubscript{$\alpha$}, respectively. Also, the use of the similarity learning CNN offers a further improvement of $0.052$ and $0.006$ mAP on the corresponding datasets. The contribution of attention mechanism to the overall performance is marginal but positive.

\begin{table}[htbp]
\caption{Impact of each network component on mAP on FIVR-5K\textsubscript{$\alpha$} and SVD\textsubscript{$\alpha$}. \textbf{MAC} stands for the features extracted from the feature extraction CNN, \textbf{W} and \textbf{A} stand for PCA whitening and attention mechanism respectively. 
}
\begin{center}
\begin{tabular}{|l||c|c|}
\hline

\textbf{Network Components} & \textbf{FIVR-5K\textsubscript{$\alpha$}} & \textbf{SVD\textsubscript{$\alpha$}}\\

\hline\hline

$\textbf{MAC}$ & $0.656$ & $0.891$\\
\hline
$\textbf{MAC} + \textbf{W}$ & $0.740$ & $0.932$\\
\hline
$\textbf{MAC} + \textbf{W} + \textbf{A}$ & $0.742$ & $0.934$\\
\hline\hline
\textbf{AuSiL} & $\textbf{0.794}$ & $\textbf{0.940}$\\
\hline

\end{tabular}
\label{parts_study}
\end{center}
\end{table}

Moreover, we investigate three different settings regarding the transfer and update of the weight parameters of the feature extraction network during training. In the settings where the network weights are updated, we do not use PCA whitening and the attention mechanism, because we encountered network collapse (the network activations were zero for any given input). Table \ref{visil_e2e} presents the results of the three variants on FIVR-5K\textsubscript{$\alpha$} and SVD\textsubscript{$\alpha$}. The settings where the parameters are transferred and not updated outperform the other two variants by a considerable margin ($0.794$ and $0.940$ mAP respectively), highlighting that transfer learning was successful. However, the poor performance of the two variants where the weights are updated is noteworthy. A possible explanation for this could be attributed to the different domains represented by the training and evaluation sets, considering that each dataset represents a domain. The network is trained on VCDB; hence, it learns the limited domain represented by this dataset. As a result, the feature extraction CNN fails to transfer knowledge and generalize to the domains of the evaluation sets, and therefore the performance drops. On the other hand, the pre-trained network is trained on AudioSet, a large-scale dataset that represents a very wide domain, and therefore the extracted knowledge can be generalized to datasets of varying domains.

\begin{table}[htbp]
\caption{mAP comparison of network variants, regarding the transfer and update of feature extraction CNN parameters during training on FIVR-5K\textsubscript{$\alpha$} and SVD\textsubscript{$\alpha$}.}
\begin{center}
\begin{tabular}{|c|c||c|c|}
\hline

\textbf{Transfer} & \textbf{Update} & \textbf{FIVR-5K\textsubscript{$\alpha$}} & \textbf{SVD\textsubscript{$\alpha$}} \\

\hline\hline

\checkmark & $\times$ & $\textbf{0.794}$ & $\textbf{0.940}$\\
\hline
\checkmark & \checkmark & $0.588$ & $0.857$\\
\hline
$\times$ & \checkmark & $0.445$ & $0.764$\\
\hline

\end{tabular}
\label{visil_e2e}
\end{center}
\end{table}

\begin{figure}[t]
    \centering
    \subfigure[FIVR-200K\textsubscript{$\alpha$}]{\includegraphics[width=7cm]{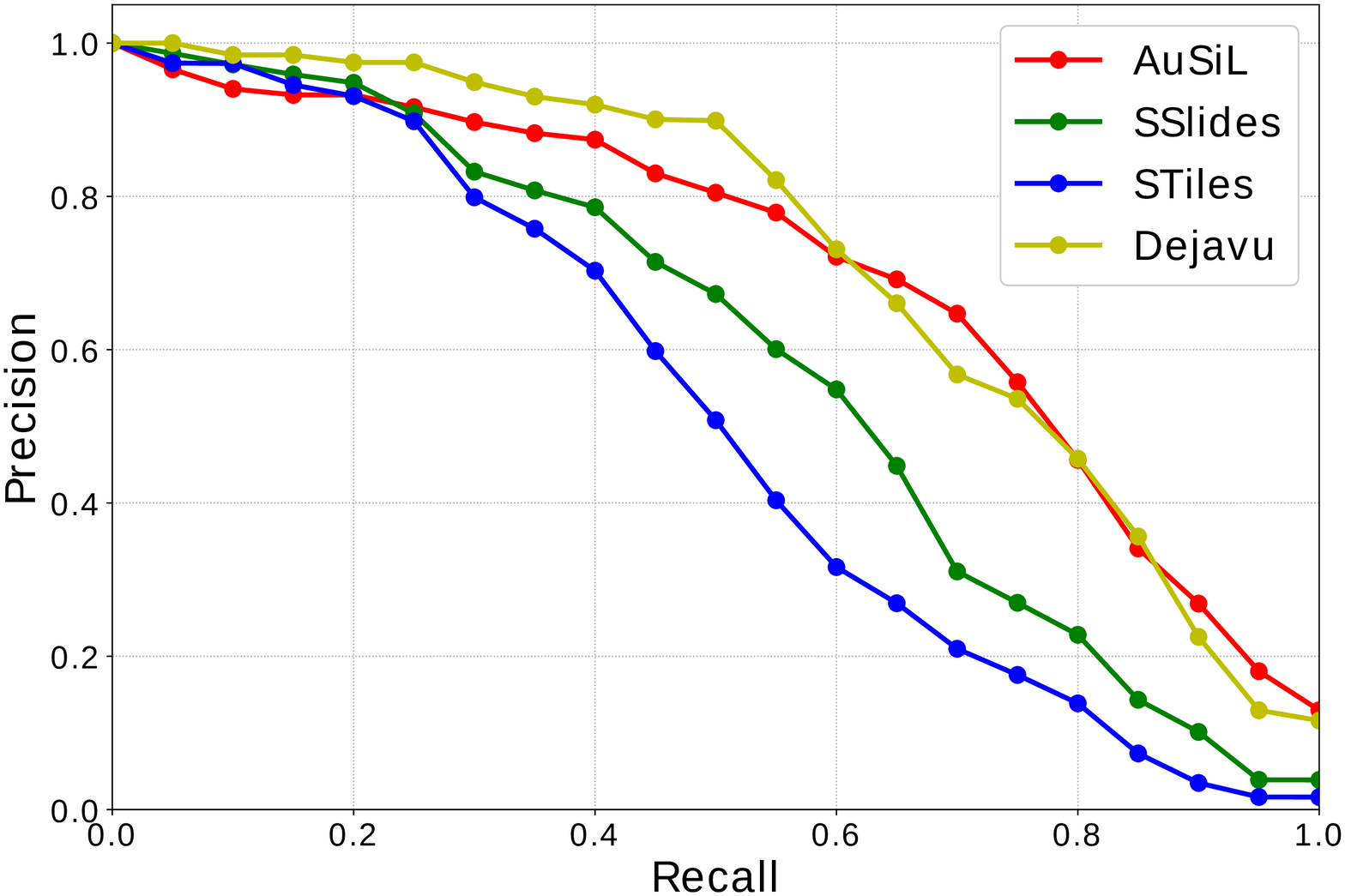}}
    \subfigure[SVD\textsubscript{$\alpha$}]{\includegraphics[width=7cm]{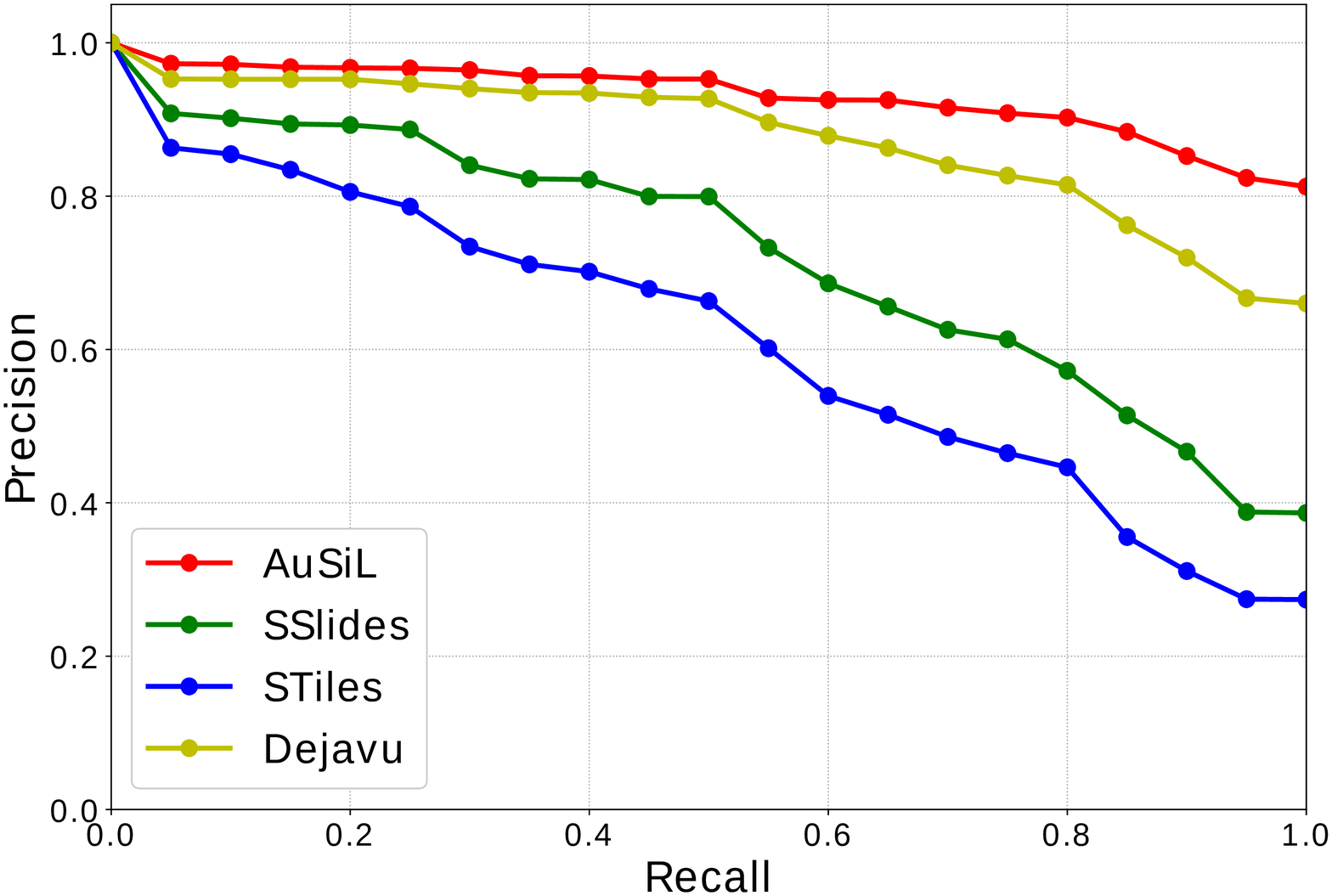}}
    \caption{Interpolated PR-curves for each approach on FIVR-200K\textsubscript{$\alpha$} and SVD\textsubscript{$\alpha$}.}
    \label{fig:pr_curve}
\end{figure}

We also investigated the impact of different values of the parameter $\gamma$, as presented in Table \ref{gamma_study}. The network performs best for $\gamma=1$, achieving $0.794$ and $0.940$ mAP on FIVR-5K\textsubscript{$\alpha$} and SVD\textsubscript{$\alpha$} respectively. For all other gamma values $g=\{0.4, 0.6, 0.8, 1.2\}$, the performance drops significantly.

\begin{table}[htbp]
    \centering
    \caption{mAP comparison for various gamma values on FIVR-5K\textsubscript{$\alpha$} and SVD\textsubscript{$\alpha$}.}
    \begin{tabular}{|l||c|c|c|c|c|}
        \hline
        gamma ($\gamma$) & 0.4 & 0.6 & 0.8 & 1.0 & 1.2 \\ \hline\hline
        \textbf{FIVR-5K\textsubscript{$\alpha$}} & $0.764$ & $0.761$ & $0.786$ & $\textbf{0.794}$ & $0.767$  \\ \hline
        \textbf{SVD\textsubscript{$\alpha$}} & $0.903$ & $0.895$ & $0.937$ & $\textbf{0.940}$ & $0.919$ \\
        \hline
    \end{tabular}
    \label{gamma_study}
\end{table}

\subsection{Comparison with the State of the art}
\label{DAVR_comparison_sec}

In this section, we benchmark our method based on the FIVR-200K\textsubscript{$\alpha$} and SVD\textsubscript{$\alpha$} datasets that we developed to simulate the DAVR problem. To evaluate the proposed AuSiL approach, we have re-implemented two state-of-the-art methods proposed for the CBCD problem \cite{ouali2014robust,  ouali2016fast}. These methods are based on binary images generated by the spectrogram of the audio signal. We will refer to them as Spectro Slides \cite{ouali2014robust} and Spectro Tiles \cite{ ouali2016fast}. Also, we compare against the publicly available open-source framework Dejavu \cite{dejavu2013}, a Shazam implementation for audio recognition.

Table \ref{DAVR_compare} illustrates the performance of the compared approaches on the two datasets. On FIVR-200K\textsubscript{$\alpha$}, the proposed approach reaches $0.701$ mAP, significantly outperforming the Spectro Slides and Tiles systems by $0.113$ mAP and $0.191$ mAP respectively; the Dejavu system surpasses the performance of AuSiL, achieving the best score of $0.726$ mAP. AuSiL achieves the best performance on SVD\textsubscript{$\alpha$} with $0.940$ mAP, outperforming all three competing methods by a significant margin, i.e., $0.066$ mAP from the second-best method, Dejavu. Looking for a reasonable explanation for the performance difference on the two evaluation datasets, we have empirically determined that a considerable amount of audio duplicates in the SVD\textsubscript{$ \alpha $} have been created using speed transformations on the audio signals. AuSiL is robust to such transformations due to its deep learning components that are trained to handle such variance. On the other hand, the other three methods rely on handcrafted methods and, therefore, can only tackle a limited range of such transformations. We set up an evaluation scheme in the next subsection in order to validate this hypothesis. Nevertheless, we experimentally found that all approaches fail to retrieve the following cases: i) the duplicate audio signal has been mixed with high volume irrelevant audio signals, e.g., speech or music, ii) the duplicate segment is too short (up to 1 second), and iii) the duplicate audio signal is very distorted.

\begin{table}[htbp]
\caption{mAP comparison of AuSiL and state-of-the-art methods on FIVR-200K\textsubscript{$\alpha$} and on SVD\textsubscript{$\alpha$}. }
\begin{center}
\begin{tabular}{|l||c|c|}
\hline

\textbf{Method} & \textbf{FIVR-200K\textsubscript{$\alpha$}} & \textbf{SVD\textsubscript{$\alpha$}}\\

\hline\hline

\textbf{Dejavu}  \cite{dejavu2013} & $\textbf{0.726}$ & $0.874$\\
\hline
\textbf{Spectro Slides} \cite{ouali2014robust} & $0.588$ & $0.716$\\
\hline
\textbf{Spectro Tiles}  \cite{ ouali2016fast} & $0.510$ & $0.605$\\
\hline\hline
\textbf{AuSiL} (ours) & $0.701$ & $\textbf{0.940}$\\
\hline

\end{tabular}
\label{DAVR_compare}
\end{center}
\end{table}

Figure \ref{fig:pr_curve} presents the Precision-Recall curves of the competing approaches on FIVR-200K\textsubscript{$\alpha$} and SVD\textsubscript{$\alpha$}, respectively. In the case of FIVR-200K\textsubscript{$\alpha $}, AuSiL's curve is below the Dejavu's curve up until 0.6 recall, outperforming or reporting similar performance on the remaining recall points. Additionally, AuSiL's curve is above those of the Spectro methods, except for the initial recall points. In the case of SVD\textsubscript{$\alpha$}, AuSiL's curve lies above all other curves with a clear margin.

\subsection{Evaluation on speed variations}
\label{sec:speed_transform}

To delve into the performance of the competing methods, we set up an evaluation scheme to benchmark the robustness of our approach to audio speed transformations. We test the FIVR-200K\textsubscript{$\alpha$} and SVD\textsubscript{$\alpha$} according to the following procedure. We first employ the dataset queries to artificially generate audio duplicates by applying speed transformations on the audio signals. We use the following factors for the generation: $\{\times0.25, \times0.5, \times0.75\}$ for slow down and $\{\times1.25, \times1.5, \times2\}$ for speed up. Then, we exclude from the datasets all videos that are originally labeled as audio duplicate, and we only consider as positives the artificially generated audio duplicates.

Our proposed method proves to be very robust on speed transformations, reaching a performance of $0.865$ and $0.923$ mAP on FIVR-200K\textsubscript{$\alpha$} and SVD\textsubscript{$\alpha$}, respectively. On the other hand, Dejavu, the best performing method on FIVR-200K\textsubscript{$\alpha$}, performs poorly, achieving only $0.443$ and $0.741$ mAP respectively. The Spectro Slides and Tiles methods do not work at all on this setup, reporting near-zero mAP. This highlights that the proposed approach tackles the limitation of the previous methods and robustly calculates the similarity between audio duplicates generated from speed transformations.

\subsection{Evaluation on visual-based tasks}
\label{visua_comp_sec}

Finally, we evaluate the performance of AuSiL and the competing approaches on the much more challenging setting of visual-based video retrieval. Although these tasks are not designed for benchmarking audio-based methods, they can still provide an indication of retrieval performance. We use the two original datasets presented in Section \ref{sec:datasets}, i.e., FIVR-200K and SVD, and also the EVVE (EVent VidEo) \cite{revaud2013event} dataset that is designed for the event-based video retrieval problem. The FIVR-200K consists of three different evaluation tasks simulating different retrieval scenarios: i) the Duplicate Scene Video Retrieval (DSVR), ii) the Complementary Scene Video Retrieval (CSVR), and iii) the Incident Scene Video Retrieval (ISVR). As expected, the performance of audio-based approaches is far worse compared with the visual-based ones, due to the fact that the visual relevance of two videos does not imply that they are also related in terms of audio.

\begin{table}[htbp]
\caption{mAP comparison of AuSiL and state-of-the-art methods on three visual-based datasets, i.e., FIVR-200K , SVD , and EVVE.}
\begin{center}
\begin{tabular}{|l||c|c|c|c|c|}
\hline
\multirow{2}{*}{\textbf{Method}} & \multicolumn{3}{c|}{\textbf{FIVR-200K }} & \multirow{2}{*}{\textbf{SVD }} & \multirow{2}{*}{\textbf{EVVE}} \\ \cline{2-4}
 & \textbf{DSVR} & \textbf{CSVR} & \textbf{ISVR} & & \\
\hline\hline
\textbf{Dejavu} \cite{dejavu2013} & $\textbf{0.352}$  & $\textbf{0.324}$ & $0.230$ & \textbf{$0.477$} & $0.160$\\ \hline
\textbf{Spectro Slides} \cite{ouali2014robust}   & $0.288$ & $0.269$ & $0.189$ & $0.406$ & $0.146$ \\ \hline
\textbf{Spectro Tiles} \cite{ ouali2016fast} & $0.249$ & $0.228$ & $0.159$ & $0.323$ & $0.144$ \\ \hline
\textbf{AuSiL} (ours) & $0.327$ & $0.310$ & $\textbf{0.232}$ & $\textbf{0.516}$ & $\textbf{0.288}$\\
\hline
\hline
\textbf{Best visual}  & ${0.892}$ & ${0.841}$ & ${0.702}$ & $0.785$ & $0.631$\\
\hline
\end{tabular}
\label{FIVR_compare}
\end{center}
\end{table}

Table \ref{FIVR_compare} presents the performance of the audio-based approaches on the FIVR-200K \cite{kordopatis2019fivr}, SVD \cite{jiang2019svd} and EVVE \cite{revaud2013event} datasets. Additionally, the table depicts the best state-of-the-art visual-based methods in each case, i.e. ViSiL \cite{kordopatis2019visil} for FIVR-200K and EVVE, and DML \cite{kordopatis2017near} for SVD. On FIVR-200K, AuSiL is outperformed by Dejavu on DSVR and CSVR tasks, but it achieves the best performance on the ISVR task with $0.232$ mAP. On SVD, AuSiL outperforms the competing audio-based approaches, achieving $0.516$ mAP and surpassing the second-best approach, Dejavu, by $0.039$ mAP. On EVVE, our approach achieves $0.288$ mAP, significantly higher than all three competing methods, with the second one reporting $0.160$ mAP. As expected, in all evaluation cases, there is a large gap in relation to the performance of the state-of-the-art visual-based approaches.

\section{Conclusions}

In this paper, we demonstrated that transfer learning and similarity learning can be effectively applied to tackle the audio-based near-duplicate video retrieval problem. In addition to achieving very competitive performance compared with three state-of-the-art approaches, the proposed architecture proved to be very robust to speed transformations of audio duplicates. A limitation of our work is that we train our network with samples derived based on the visual duplicity of videos, and without explicitly knowing if they are actually audio duplicates. Thus, employing a training set with proper audio annotation could further boost retrieval performance. For future work, we plan to examine different feature extraction methods with different network architectures, tailored for the application of the proposed scheme to similar tasks, e.g., cover song detection. Also, we will investigate ways of reducing the computational complexity of the proposed method.

\section*{Acknowledgments}
This work has been supported by the WeVerify project, partially funded by the European Commission under contract number 825297, and the News.vid.io project, funded by the Google DNI under contract number 555950.

\bibliographystyle{IEEEtran}
\bibliography{ref}

\begin{thebibliography}{10}
\providecommand{\url}[1]{#1}
\csname url@samestyle\endcsname
\providecommand{\newblock}{\relax}
\providecommand{\bibinfo}[2]{#2}
\providecommand{\BIBentrySTDinterwordspacing}{\spaceskip=0pt\relax}
\providecommand{\BIBentryALTinterwordstretchfactor}{4}
\providecommand{\BIBentryALTinterwordspacing}{\spaceskip=\fontdimen2\font plus
\BIBentryALTinterwordstretchfactor\fontdimen3\font minus
  \fontdimen4\font\relax}
\providecommand{\BIBforeignlanguage}[2]{{%
\expandafter\ifx\csname l@#1\endcsname\relax
\typeout{** WARNING: IEEEtran.bst: No hyphenation pattern has been}%
\typeout{** loaded for the language `#1'. Using the pattern for}%
\typeout{** the default language instead.}%
\else
\language=\csname l@#1\endcsname
\fi
#2}}
\providecommand{\BIBdecl}{\relax}
\BIBdecl

\bibitem{deng2009imagenet}
J.~Deng, W.~Dong, R.~Socher, L.-J. Li, K.~Li, and L.~Fei-Fei, ``Imagenet: A
  large-scale hierarchical image database,'' in \emph{IEEE conference on
  Computer Vision and Pattern Recognition}, 2009.

\bibitem{kumar2018knowledge}
A.~Kumar, M.~Khadkevich, and C.~F{\"u}gen, ``Knowledge transfer from weakly
  labeled audio using convolutional neural network for sound events and
  scenes,'' in \emph{IEEE International Conference on Acoustics, Speech and
  Signal Processing}, 2018.

\bibitem{audioset}
J.~F. Gemmeke, D.~P.~W. Ellis, D.~Freedman, A.~Jansen, W.~Lawrence, R.~C.
  Moore, M.~Plakal, and M.~Ritter, ``Audio set: An ontology and human-labeled
  dataset for audio events,'' in \emph{IEEE International Conference on
  Acoustics, Speech and Signal Processing}, 2017.

\bibitem{kordopatis2019visil}
G.~Kordopatis-Zilos, S.~Papadopoulos, I.~Patras, and I.~Kompatsiaris,
  ``{ViSiL}: Fine-grained spatio-temporal video similarity learning,'' in
  \emph{IEEE International Conference on Computer Vision}, 2019.

\bibitem{kordopatis2019fivr}
------, ``{FIVR}: Fine-grained incident video retrieval,'' \emph{IEEE
  Transactions on Multimedia}, 2019.

\bibitem{jiang2019svd}
Q.-Y. Jiang, Y.~He, G.~Li, J.~Lin, L.~Li, and W.-J. Li, ``{SVD}: A large-scale
  short video dataset for near-duplicate video retrieval,'' in \emph{IEEE
  International Conference on Computer Vision}, 2019.

\bibitem{roopalakshmi2011novel}
R.~Roopalakshmi and G.~R.~M. Reddy, ``A novel approach to video copy detection
  using audio fingerprints and pca,'' \emph{Procedia Computer Science}, 2011.

\bibitem{jegou2012babaz}
H.~J{\'e}gou, J.~Delhumeau, J.~Yuan, G.~Gravier, and P.~Gros, ``Babaz: a large
  scale audio search system for video copy detection,'' in \emph{IEEE
  International Conference on Acoustics, Speech and Signal Processing}, 2012.

\bibitem{haitsma2002highly}
J.~Haitsma and T.~Kalker, ``A highly robust audio fingerprinting system.'' in
  \emph{International Conference on Music Information Retrieval}, 2002.

\bibitem{saracoglu2009content}
A.~Saracoglu, E.~Esen, T.~K. Ates, B.~O. Acar, U.~Zubari, E.~C. Ozan, E.~Ozalp,
  A.~A. Alatan, and T.~Ciloglu, ``Content based copy detection with coarse
  audio-visual fingerprints,'' in \emph{International Workshop on Content-Based
  Multimedia Indexing}, 2009.

\bibitem{wang2012contented}
L.~Wang, Y.~Dong, H.~Bai, J.~Zhang, C.~Huang, and W.~Liu, ``Contented-based
  large scale web audio copy detection,'' in \emph{IEEE International
  Conference on Multimedia and Expo}, 2012.

\bibitem{wang2003industrial}
A.~Wang, ``An industrial strength audio search algorithm.'' in
  \emph{International Conference on Music Information Retrieval}, 2003.

\bibitem{anguera2012mask}
X.~Anguera, A.~Garzon, and T.~Adamek, ``Mask: Robust local features for audio
  fingerprinting,'' in \emph{IEEE International Conference on Multimedia and
  Expo}, 2012.

\bibitem{ouali2014robust}
C.~Ouali, P.~Dumouchel, and V.~Gupta, ``A robust audio fingerprinting method
  for content-based copy detection,'' in \emph{International Workshop on
  Content-Based Multimedia Indexing (CBMI)}, 2014.

\bibitem{ouali2015efficient}
------, ``Efficient spectrogram-based binary image feature for audio copy
  detection,'' in \emph{IEEE International Conference on Acoustics, Speech and
  Signal Processing}, 2015.

\bibitem{ouali2016fast}
------, ``Fast audio fingerprinting system using gpu and a clustering-based
  technique,'' \emph{IEEE/ACM Transactions on Audio, Speech, and Language
  Processing}, 2016.

\bibitem{dejavu2013}
worldveil, ``Audio fingerprinting and recognition in python,''
  \url{https://github.com/worldveil/dejavu}, 2013.

\bibitem{tolias2015particular}
G.~Tolias, R.~Sicre, and H.~J{\'e}gou, ``Particular object retrieval with
  integral max-pooling of cnn activations,'' \emph{arXiv:1511.05879}, 2015.

\bibitem{kordopatis2017a}
G.~Kordopatis-Zilos, S.~Papadopoulos, I.~Patras, and Y.~Kompatsiaris,
  ``Near-duplicate video retrieval by aggregating intermediate cnn layers,'' in
  \emph{International conference on multimedia modeling}, 2017.

\bibitem{jegou2012negative}
H.~J{\'e}gou and O.~Chum, ``Negative evidences and co-occurences in image
  retrieval: The benefit of pca and whitening,'' in \emph{European Conference
  on Computer Vision}, 2012.

\bibitem{yang2016hierarchical}
Z.~Yang, D.~Yang, C.~Dyer, X.~He, A.~Smola, and E.~Hovy, ``Hierarchical
  attention networks for document classification,'' in \emph{conference of the
  North American chapter of the association for computational linguistics:
  human language technologies}, 2016.

\bibitem{schroff2015facenet}
F.~Schroff, D.~Kalenichenko, and J.~Philbin, ``Facenet: A unified embedding for
  face recognition and clustering,'' in \emph{Proceedings of the IEEE
  conference on computer vision and pattern recognition}, 2015.

\bibitem{jiang2014vcdb}
Y.-G. Jiang, Y.~Jiang, and J.~Wang, ``{VCDB}: a large-scale database for
  partial copy detection in videos,'' in \emph{European Conference on Computer
  Vision}, 2014.

\bibitem{kingma2014adam}
D.~P. Kingma and J.~Ba, ``Adam: A method for stochastic optimization,''
  \emph{arXiv preprint arXiv:1412.6980}, 2014.

\bibitem{revaud2013event}
J.~Revaud, M.~Douze, C.~Schmid, and H.~J{\'e}gou, ``Event retrieval in large
  video collections with circulant temporal encoding,'' in \emph{IEEE
  conference on Computer Vision and Pattern Recognition}, 2013.

\bibitem{kordopatis2017near}
G.~Kordopatis-Zilos, S.~Papadopoulos, I.~Patras, and Y.~Kompatsiaris,
  ``Near-duplicate video retrieval with deep metric learning,'' in \emph{IEEE
  International Conference on Computer Vision Workshops}, 2017.

\end{thebibliography}


\end{document}